# Element Abundances at High Redshifts:
# The N/O Ratio at Low Metallicity

Keith Lipman[1], Max Pettini[2] and Richard W. Hunstead[3]

[1] Institute of Astronomy, Madingley Road, Cambridge, CB3 0HA, UK
(kl@mail.ast.cam.ac.uk)

[2] Royal Greenwich Observatory, Madingley Road, Cambridge, CB3 0EZ, UK
(pettini@mail.ast.cam.ac.uk)

[3] School of Physics, University of Sydney, NSW 2006, Australia
(rwh@astrop.physics.su.oz.au)

ABSTRACT

Our knowledge of galactic chemical evolution is currently limited to observations of Milky Way stars and H II regions of nearby galaxies. Damped Lyman $\alpha$ systems offer a new approach for tracking the evolution of normal galaxies from early epochs to the present day. Here we report the first measurements of nitrogen abundances in galaxies with less than 1/100 of solar metallicity, a range unexplored by previous observations.

## 1. INTRODUCTION

High resolution spectroscopy of absorbers along QSO sightlines allows us to study element abundances in high redshift galaxies with sufficient accuracy to test models of chemical evolution. The damped Lyman $\alpha$ population, thought to represent the progenitors of present day galaxies (Wolfe, this volume), is particularly well suited to such studies: the large neutral hydrogen column densities not only make accessible absorption lines of weak transitions, but also ensure that the gas is optically thick, largely removing the need for ionization corrections. It is of particular interest that many damped Lyman $\alpha$ systems are very metal-poor, offering us a view of the earliest stages of chemical enrichment in galaxies.

Traditionally, galactic chemical evolution studies have relied on observations of Galactic stars and of bright H II regions in nearby galaxies. Both classes of object have revealed much about the process of metal enrichment; however, each has its limitations. The usefulness of the H II regions is restricted by their abundances; few are known with $Z < 1/10\ Z_\odot$, and none with $Z \lesssim 1/50\ Z_\odot$. Furthermore, local pollution from evolved stars such as Wolf-Rayets can confuse the interpretation, especially at low metallicities. While old stellar populations do extend to very low abundances, observations of some elements – such as nitrogen – are particularly difficult in metal-poor stars. A further point is that existing abundance measurements in both stars and H II regions refer exclusively to nearby galaxies; at present we have no way of observing these objects in the more distant universe.

High resolution spectroscopy of damped Lyman $\alpha$ systems therefore has much to offer. We have started a program to investigate the abundance of nitrogen in metal-poor damped Lyman $\alpha$ systems, extending the studies of H II regions and stars to lower metallicities than previously attainable. This approach promises to cast light on the nucleosynthetic origin of N, which is not yet fully understood.

## 2. OBSERVATIONS

We have analyzed two damped Lyman $\alpha$ systems so far in this project, both chosen for their low metallicities. The $z_{\rm abs} = 2.53788$ system towards Q2344+124 ($z_{\rm em} = 2.779$) was observed with the William Herschel telescope on La Palma, while the $z_{\rm abs} = 2.27936$ absorber towards Q2348−147 ($z_{\rm em} = 2.940$) was observed at the Anglo-Australian telescope. In both cases we used echelle spectrographs and photon-counting detectors to achieve a resolution of $R \simeq 45\,000$, corresponding to $FWHM \simeq 7$ km s$^{-1}$. A full description of the observations is given by Pettini, Lipman and Hunstead (1995) and Lipman and Pettini (1995).

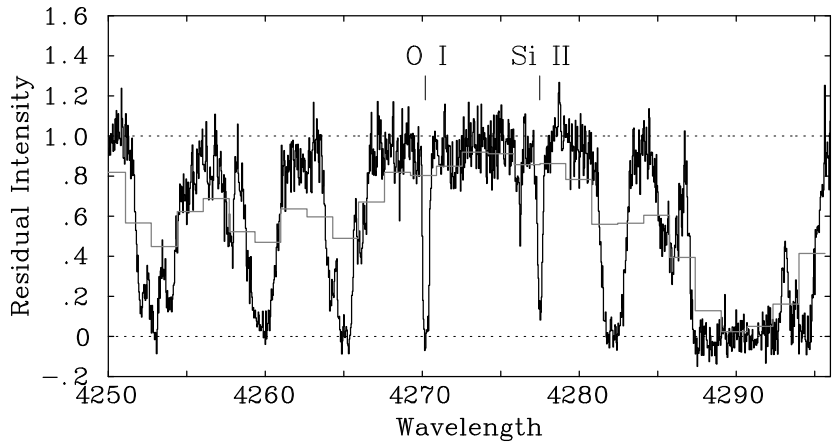

Figure 1: Portion of the AAT IPCS spectrum of Q2348−147 encompassing the O I $\lambda 1302$ and Si II $\lambda 1304$ transitions in the $z_{\rm abs} = 2.27936$ damped Lyman $\alpha$ system. The resolving power is $R \simeq 45\,000$; the rest-frame equivalent widths of the O I and Si II lines are 146 mÅ and 102 mÅ respectively. The grey line shows how the same data would appear if recorded with the coarser $R = 1300$ of the $HST$ observations by Reimers et al. (1992). Narrow lines would be lost in the noise and only strong, saturated features – quite unsuitable for abundance measurements – would be discerned.

### 3. ELEMENT ABUNDANCES

A portion of the spectrum of Q2348−147 is shown in Figure 1. The velocity structure of the $z_{\rm abs} = 2.27936$ damped system is particularly simple, apparently consisting of a single absorption component with Doppler parameter $b \simeq 10$ km s$^{-1}$. Column densities $N$ of species which are the major ion stages in H I regions were deduced by fitting theoretical absorption profiles to the metal lines in the manner described by Pettini, Lipman and Hunstead (1995). Errors were estimated by exploring the $b - N$ parameter space to determine the range of values of $N$ compatible with the shape of the observed profiles and their equivalent widths (to within $1\sigma$).

Dividing by the neutral hydrogen column density $N({\rm H\,I}) = (3.7 \pm 0.7) \times 10^{20}$ cm$^{-2}$, implied by the damping wings of the Lyman $\alpha$ line, then leads to the element abundances shown in Figure 2.

The absorption lines of C II and O I are saturated; in such cases it is not possible to limit the corresponding abundances to better than 3 orders of magnitude, despite the high resolution of our echelle spectra. The problem is exacerbated when dealing with low resolution data, such as those of Reimers et al. (1992), since *only* saturated lines would then be detected (see Figure 1). For this reason, abundance measurements from low resolution observations must be viewed with caution.

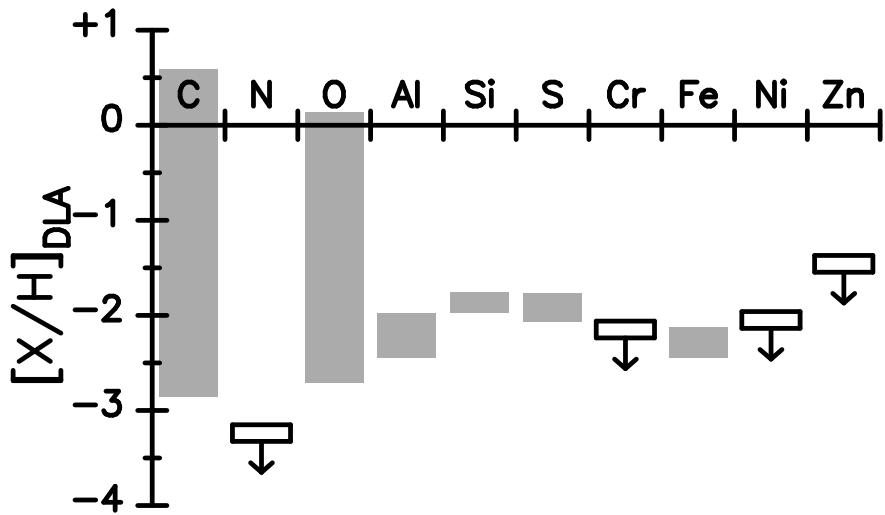

Figure 2: Element abundances in the $z_{\rm abs} = 2.27936$ damped Lyman $\alpha$ system towards Q2348−147. Values are plotted on a logarithmic scale relative to solar abundances (Anders and Grevesse 1989). Grey vertical bars indicate positive detections, the length of the bar reflecting the allowed range of values. Open boxes show upper limits; the upper and lower edges of the boxes represent $3\sigma$ and $2\sigma$ limits respectively.

Figure 2 shows that most elements in the galaxy giving rise to the $z_{\rm abs} = 2.27936$ damped system are underabundant by about two orders of magnitude relative to the Sun. The fraction of heavy elements present in dust grains is also significantly reduced compared with the local interstellar medium. This is readily realised when we consider that elements such as Fe, Al and S, which locally exhibit widely differing degrees of dust depletions – by up to two orders of magnitude – are found in roughly solar relative proportions in this high-redshift galaxy. A striking feature of Figure 2 is the underabundance of N relative to other elements. This is not a consequence of line misidentification or blending, since the N limit is imposed by a non-detection.

As discussed below, it is useful to compare the N abundance with that of O. To circumvent the large uncertainty in [O/H], we have assumed the O abundance to be equal to that of S in the case of Q2348−147, and that of Si in Q2344+124. These are plausible assumptions since all three elements are created via the $\alpha$ process, and their relative abundances are observed to be roughly constant over the full range of metallicities sampled by H II regions and stellar studies (e.g. Skillman and Kennicutt 1993; Garnett and Kennicutt 1994; McWilliam et al. 1995). Furthermore, we do not expect dust depletions to alter significantly the abundance pattern in either absorption system.

# 4. THE NITROGEN TO OXYGEN RATIO

Why N/O? Perhaps originally this ratio was studied for no better reason than that both elements are easily observable in H II region emission-line spectra. However, the choice does serve a useful purpose. The nucleosynthetic origin of O is reasonably well understood, while that of N is not; the relative abundances of these two elements can be used to test current ideas on the mechanisms responsible for the formation of N.

*Primary* production of an element involves its formation from seed nuclei which are produced *in situ* within the star. Primary production does not require the presence of pre-existing nuclei and is therefore independent of the initial metallicity of the star. O is a good example of a primary element, and its production is dominated by supernova explosions of massive stars. The major producers of primary N are thought to be stars of intermediate mass during the third dredge-up phase (Renzini and Voli 1981).

*Secondary* N, on the other hand, is synthesised from pre-existing seed nuclei. Such a process is thought to be possible in most stars, except those of low mass in which the core temperature is too low to initiate the CNO cycle. If N is produced in this manner, its yield is expected to be proportional to the initial metallicity of the star.

Figure 3 shows [N/O] as a function of [O/H] for a variety of observations. Also shown are the expectations for secondary only, and combined primary and secondary origin of N (Vila-Costas and Edmunds 1993). Primary production dominates at low values of [O/H], but secondary nitrogen dominates at higher abundances. Our measurements of [N/O] in the damped systems lie in a distinct region of the plot, within the bounds predicted for primary and secondary production, but at lower metallicities than those sampled up to now.

There are several effects to consider when interpreting Figure 3. While H II regions are rich in recent stellar and supernova ejecta, the QSO sightlines probe neutral gas, not necessarily close to regions of active star formation. Therefore, the evolution of the N/O ratio in the population of galaxies traced by the damped systems depends on the rate at which freshly synthesized elements mix with the more generally distributed interstellar medium. Furthermore, the position of each point not only depends on the production mechanism of N, but also on the previous star formation history. For example, were we to observe an H II region soon after a starburst, we would expect that O, produced from short-lived massive stars, would be overabundant relative to N which, being produced mostly in stars of intermediate mass is released back into the ISM over longer timescales (e.g. Edmunds and Pagel 1978). If star formation takes place in isolated starbursts separated by quiescent periods, this delayed-release scenario predicts that the scatter in [N/O] should be greater at lower metallicities (Garnett 1990; Olofsson 1995). Our survey will be able to establish if such a scatter exists among high-redshift damped Lyman $\alpha$ galaxies.

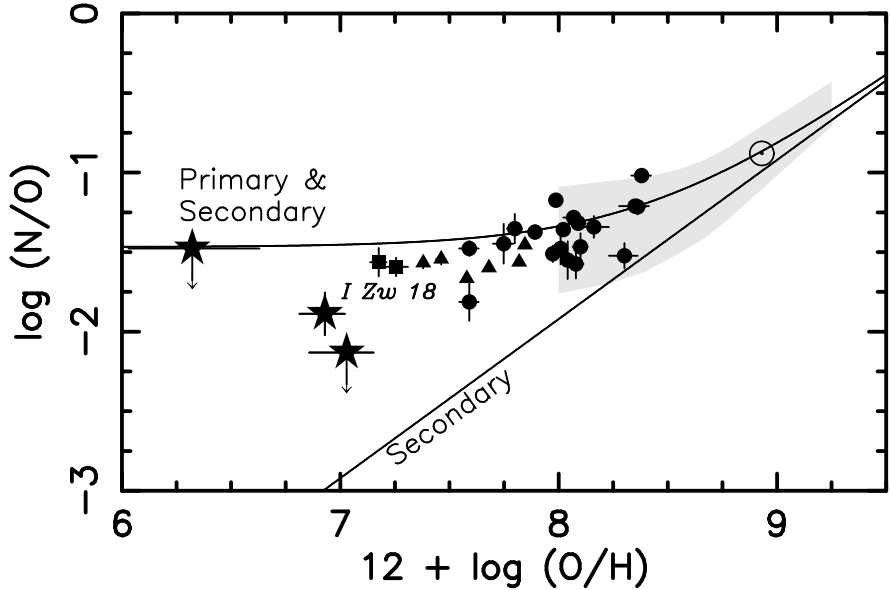

Figure 3: The N/O ratio for three damped Lyman $\alpha$ systems (star symbols) are compared with values measured in H II regions in spiral galaxies (shaded area) and dwarf star-forming galaxies (filled symbols). H II region data are from the extensive compilation by Vila-Costas and Edmunds (1993); the dwarf galaxies selected are those whose spectra Pagel et al. (1992 – circles) and Izotov et al. (1994 – triangles) considered to be free of contamination by Wolf-Rayet features. The two squares refer to the two H II regions in I Zw 18 observed by Skillman and Kennicutt (1993). The damped Lyman $\alpha$ systems are, from the left, HS 1946+7658 ($z_{abs} = 2.8443$; Fan and Tytler 1994), Q2344+124 ($z_{abs} = 2.53788$; Lipman and Pettini 1995) and Q2348−147 ($z_{abs} = 2.27936$; this paper). Symbols with downward-pointing arrows indicate upper limits to [N/O]. The Sun symbol shows the solar system abundances (Anders and Grevesse 1989). Also shown are the predictions for a purely secondary and a primary + secondary origin of N, reproduced from Vila-Costas and Edmunds (1993).

In conclusion, even the preliminary observations presented here provide a clear demonstration of the high potential of damped Lyman $\alpha$ systems, not only for understanding the nucleosynthetic origin of N, but also for studying many other aspects of galactic chemical evolution. We can expect an exciting future in this field now that we have entered the era of 10m-class telescopes.